\documentstyle[aps,epsfig,multicol]{revtex}

\begin{document}

\title{Evidence for a generic novel quantum transition in high-$T_c$ cuprates }
\author{C. Panagopoulos$^1$, J.L.Tallon$^2$, B.D. Rainford$^3$, T. Xiang$^4$, J.R.
Cooper$^1$ and C.A. Scott$^3$ }
\address{$^1$ Cavendish Laboratory and IRC in Superconductivity, University of
Cambridge, Cambridge CB3 0HE, United Kingdom}
\address{$^2$ New Zealand Institute for Industrial Research, P.O. Box 31310, Lower
Hutt, New Zealand}
\address{$^3$ Department of Physics and Astronomy, University of Southampton,
Southampton S017 1BJ, United Kingdom}
\address{$^4$ Institute of Theoretical Physics, Academia Sinica, P.O. Box 2735, Beijing
100080, Peoples Republic of China}

\date{\today}

\maketitle

\begin{abstract}
We study the low-energy spin fluctuations and superfluid density of a series of pure and
Zn-substituted high-$T_c$ superconductors (HTS) using the muon spin relaxation and
$ac$-susceptibility techniques.
At a critical doping state, $p_c$, we find (i) simultaneous
abrupt changes in the magnetic spectrum in the superconducting ground state and (ii)
that the slowing down of spin fluctuations becomes singular at $T=0$. These results
provide experimental evidence for a novel quantum transition that separates the
superconducting phase diagram of HTS into two distinct ground states.
\end{abstract}

\begin{multicols}{2}

Quantum phase transitions occur at zero-temperature at a critical electron density
separating distinct ground states. Near a quantum critical point electrons in metals are
highly correlated and the diverging fluctuations may induce unconventional
superconductivity \cite{SC,SY1,CC,CV1,DP,Si}. For example,
in certain heavy fermion
compounds a ``bubble'' of superconductivity occurs around the quantum
critical point at
which itinerant antiferromagnetism is suppressed by applied pressure \cite{Mathur}. The
search for an underlying quantum phase transition in HTS is motivated by the potential
for quantum fluctuations to bind electronic carriers into
superconducting Cooper pairs
and also to cause the celebrated linear temperature dependence of their electrical
resistivity \cite{SC,SY1,CC,CV1,DP,Si,TallonLoram}. HTS exhibit a common generic
phase diagram in which the superconducting transition temperature, $T_c$ rises to a
maximum at an optimal doping of approximately 0.16 holes per planar copper atom and
then falls to zero on the overdoped side. In addition the underdoped normal state exhibits
correlations, which introduce a gap in the density of states that strongly affects all
physical properties. There is no phase transition associated with the opening of this gap
and so it is called a pseudogap. Analysis of specific heat data, for example, suggests that
the pseudogap energy decreases with doping and falls to zero at a critical doping of
$p_c$ $\simeq$ 0.19, just beyond optimal doping \cite{TallonLoram,Loram}, a
behaviour rather analogous to the quantum-critical heavy-fermion materials
\cite{Mathur}.

Many fundamental physical quantities such as the superconducting condensation energy
\cite{TallonLoram,Loram}, the superfluid density \cite{LSC}, and the quasiparticle
lifetime \cite{TallonLoram,Feng}, show abrupt changes as $p$ $\rightarrow$ $p_c$.
While compelling in their totality \cite{TallonLoram,Loram}, none of the results can be
considered as evidence of a quantum transition. In particular there is no evidence for an
associated order parameter and slowing down of the relevant fluctuations. With this in
mind we examined the evolution with doping of the low-energy spin fluctuation spectrum
using muon spin relaxation ($\mu$SR) combined with low-field $ac$-susceptibility
measurements of the superfluid density.

The samples studied were: (i) La$_{2-x}$Sr$_x$Cu$_{1-y}$Zn$_y$O$_4$ (LSCO)
($x$=0.03-0.24 and $y$=0, 0.01 and 0.02). Samples were synthesised using solid-state
reaction and where necessary followed by quenching and subsequent oxygenation. They
were characterised by powder x-ray diffraction as well as extensive transport and
thermodynamic measurements $e.g.$ refs  \cite{Loram,LSC,CP4} and found to be phase
pure. Their $T_c$ values and lattice parameters were also in good agreement with
published data, where available \cite{PGR,Kastner}. (ii) Bi$_{2.1}$Sr$_{1.9}$Ca$_{1-
x}$Y$_x$Cu$_2$O$_{8+y}$ (Bi-2212) ($x$=0, 0.3, 0.5 and appropriate values of $y$
to achieve the desired carrier concentration). Underdoped samples were prepared by
deoxygenation and the samples were fully characterised using also thermoelectric power
to determine the doping state. (Note that deoxygenation actually removes disorder in this
system.) We note that in LSCO $x=p$ and to avoid confusion in the rest of the paper we
refer to the carrier concentration as $p$.

Zero-field (ZF) and transverse-field (TF) $\mu$SR studies were performed at the pulsed
muon source, ISIS Facility, Rutherford Appleton Laboratory. Spectra were collected
down to as low as 40mK thus allowing the temperature dependence of slow spin
fluctuations to be studied to high doping. In a $\mu$SR experiment, $100\%$ spin-
polarised positive muons implanted into a specimen precess in their local magnetic
environment. Random spin fluctuations will depolarise the muons provided they do not
fluctuate much faster than the muon precession. The muon decays with a life-time
$2.2\mu s$, emitting a positron preferentially in the direction of the muon spin at the
time of decay. By accumulating time histograms of such positrons one may deduce the
muon depolarisation rate as a function of time after implantation. The muon is expected
to reside at the most electronegative site of the lattice. In both HTS families studied here
it is the O$^{-2}$ nearest to the planes \cite{muonsite}
so the results reported here are
dominated by the magnetic correlations in the CuO$_2$ planes. As we show below this
is confirmed by results in samples doped with Zn, which substitutes for Cu in the
CuO$_2$ planes.

The superfluid density, $\lambda_{ab}^{-2}$, results shown here are for pure La$_{2-
x}$Sr$_x$CuO$_4$ ($y=0$) and were determined from measurements of the in-plane
magnetic penetration depth $\lambda_{ab}$ using the low-field $ac$-susceptibility
technique at 1G and 333Hz for grain-aligned powders \cite{LSC,CP4,CP1}.  In total, 16
samples were investigated for each doping content. The values of $\lambda_{ab}(0)^{-
2}$ for samples with $p\geq 0.15$ were also confirmed by standard TF $\mu$SR
experiments performed on unaligned powders at 400G \cite{LSC}.

Figure 1 shows the typical time dependence of the ZF muon asymmetry
\cite{Kiefl,CN,YU,BDR} for pure La$_{2-x}$Sr$_x$CuO$_4$ with $x=p$=0.08
($T_c$=21K) at (a) low and (b) high temperatures. In all samples the high-temperature
form Fig. 1b of the depolarisation is Gaussian and temperature independent, consistent
with dipolar interactions between the muons and their near-neighbour nuclear moments.
This was verified by applying a 50G longitudinal field, which completely suppressed the
depolarisation. Here the electronic spins in the CuO$_2$ planes fluctuate so fast that they
do not affect the muon polarisation. At low enough temperatures Fig. 1a, typical of other
spin glass systems \cite{muonsite,Kiefl,CN,YU,BDR}, there is a fast relaxation due to a
static distribution of random local fields, followed by a long-time tail with a slower
relaxation resulting from remnant dynamical processes within the spin glass. At very low
temperatures oscillations in the asymmetry were observed for $p\leq 0.08$ and by
decoupling experiments in a longitudinal field we confirmed the static nature of the
magnetic ground state. For $p>0.08$ oscillations were not observed and, as discussed
below, the data were better represented by an exponential relaxation indicating either a
very strongly disordered static field distribution or rapid fluctuations. Between the high
and low temperature limits the spin correlations slow down through the experimental
$\mu$SR time window and modify the depolarisation process in a distinctive fashion.

To study the doping dependence of this slowing down we determine two characteristic
temperatures. (i) The temperature, $T_f$, where the spin correlations first enter the
$\mu$SR time window i.e., where the muon asymmetry first deviates from Gaussian
behaviour and (ii) the temperature, $T_g$, at which these correlations freeze into a glassy
state thus causing an initial rapid decay in the asymmetry. $\mu$SR is sensitive to spin
fluctuations within a time window of $10^{-9}$s to $10^{-6}$s \cite{YU} and we may
therefore associate $T_f$ and $T_g$, respectively, with these lower and upper
thresholds. In general the relaxation data may be fitted to the form $G_z(t)=A_1 exp( -
\gamma_1 t) +A_2 exp(- (\gamma_2 t)^\beta) +A_3$ where the first term is the fast
relaxation in the glassy state (i.e., at higher temperatures $A_1=0$),
the second stretched-exponential term is the slower dynamical term and $A_3$ accounts for a
small time-independent background arising from muons stopping in the silver backing
plate. As in other spin glass systems,
in the high-temperature Gaussian limit $\beta$=2.0
\cite{muonsite,YU,BDR}. Consequently, any departure below $\beta$=2.0$\pm0.06$ is
taken as the onset temperature, $T_f$, at which spin fluctuations slow down sufficiently
to enter the time scale of the muon probe ($10^{-9}$s). A typical temperature
dependence for $\beta$ for Sr=0.08 is shown in Fig. 2a. (As a further check on the
assignment of $T_f$ we fitted the high-temperature data to the full Kubo-Toyabe
function $G_z(t)=A_1exp(- \alpha t^2)exp(- \gamma t)+A_2$ and values of the
relaxation rate $\gamma$ are plotted in Fig. 2b. $\gamma$ is found to rise from zero at
the same temperature at which the exponent $\beta$ departs from 2, indicating the
entrance of the spin correlations into the experimental time window.) At low
temperatures the exponent $\beta$ falls rapidly towards the value 0.5. We identify $T_g$
as the temperature at which $\beta$=0.5$\pm0.06$ \cite{muonsite,YU,BDR}.
This ``root exponential'' form for the relaxation function is a common feature of spin glasses, and in
the present samples the temperature $T_g$ coincided with a maximum in the
longitudinal relaxation rate $\gamma$ (see Fig. 2b) and the appearance of the fast
relaxation. Other methods of analysis may still be possible and a different choice might
affect the magnitude of the $T_g$ values but not the trends. Our values for $T_g$ agree
with published data, where available \cite{Kastner,CN,Singer}.

We first discuss the data for pure LSCO (i.e., $y=0$ in Fig. 3). Values of $T_g$ and
$T_f$ summarised in Fig. 3 indicate that the spin-glass phase persists beyond $p=0.125$.
In fact the onset of the spin glass phase for $p=0.125$ occurs at a higher temperature
than that for $p=0.10$. This may be due to the formation of strongly-correlated
antiferromagnetic stripe domains in this range of doping \cite{Kastner,Kiv,Zaanen}. For
$p=0.15$ and 0.17, $T_g$ becomes very small (less than 45mK) and $T_f$ is approximately 8K
and 2K, respectively. For $p\geq 0.20$, there are no changes in the depolarisation
function to the lowest temperature measured (40mK).

Figure 3 clearly shows that although the freezing of spins occurs at very low
temperatures, low-frequency spin correlations enter the experimental time window at
significantly higher temperatures. Both $T_g$ and $T_f$ are found to decrease with
increasing doping and tend to zero at $p\simeq 0.19$. Their behaviour resembles that of
the pseudogap \cite{TallonLoram,Loram} which vanishes at the same doping. The LSCO
results are reproduced in the Bi-2212 system (Fig. 3), which shows precisely the same
trend with $T_g$ and $T_f$ $\rightarrow$ 0 as $p$ $\rightarrow$ 0.19. This observation
is not to discount any effects of stripes on our LSCO data.  Indeed such effects are clearly
seen in the figure where the anomaly seen in LSCO at the hole concentration of 0.125 is
clearly absent or diminished in the Bi-2212 data. We also know from a couple of data
points \cite{CP248} that the YBa$_2$Cu$_3$O$_{7-\delta}$ system trends in the same
way as LSCO and Bi-2212. Similar slowing down of spin fluctuations has also been
observed in the most ordered of all HTS namely the underdoped YBa$_2$Cu$_4$O$_8$
\cite{CP248}. Furthermore, the doping dependence of $T_g$ seen here has also been
found in Y$_{1-y}$Ca$_y$Ba$_2$Cu$_3$O$_{6.02}$ up to 0.09 holes per planar
copper atom \cite{CN}. These observations indicate that the behaviour shown in Fig. 3 is
common to all HTS. They also indicate that the observed trends are not a consequence of
a structural transition or inhomogeneity peculiar to a specific HTS family.

Earlier spectroscopic studies have shown that substitution with Zn slows down the spin
correlations \cite{Kastner,Julien,Kimura}. This, as depicted in Fig. 3, enhances the muon
depolarisation rate at low temperatures and causes an increase in both $T_g$ and $T_f$.
The striking result which Fig. 3 summarises is the apparent convergence of both
$T_{g}(p)$ and $T_{f}(p)$ to zero, for all Zn concentrations, at the critical doping
$p_{c}\simeq 0.19$. While this effect is not so obvious for the pure LSCO samples it is
very clear in the two Zn-substituted series. The fact that $T_{f}(p)$ $\rightarrow$ 0 as
$p$ $\rightarrow$ $p_{c}$ for all Zn concentrations suggests that spin correlations
within the upper $\mu$SR time threshold of $10^{-9}$s die out abruptly beyond $p_{c}$
leaving only short-lived fluctuations (or none at all) beyond $p_{c}$. The fact that $T_f$
and $T_g$ both vanish as $p$ $\rightarrow$ $p_c$ implies that the rate of slowing down
$diverges$ at $p_c$, (i.e. the characteristic time changes from $10^{-9}$s to $10^{-6}$s in
smaller and smaller temperature intervals as $p_c$ is approached). In the absence of
evidence for long-range order in the normal-state, the present observations indicate the
existence of a quantum glass transition at $p_c$. We note that if the observed quantum
glass transition is a conventional spin glass transition, then the glass transition at $T=0$
is indeed a quantum critical point \cite{SY2}. The present behaviour could alternatively
be driven by the existence of a quantum metal-insulator transition as has been observed
in La$_{2-x}$Sr$_x$CuO$_4$ near $Sr$=0.18 \cite{Boebinger}. For $p>p_c$ spin-flip
scattering associated with mobile holes could reduce the lifetime of the spins sufficiently
to prevent freezing. Either way, the present results demonstrate the disappearance of
short-range magnetic order at $p_c$ and a clear link between a quantum transition, the
essential physics of superconductivity and the pseudogap in HTS.

This link is further underscored by our detailed measurements of the doping dependence
of $\lambda_{ab}^{-2}(0)$ for pure LSCO shown in Fig. 4. The superfluid density
remains constant above $p_c$ but falls abruptly below $p_c$. A similar doping
dependence of the superfluid density has also been observed in Bi-2212 \cite{Anukool}.
This confirms again that this phase behaviour is generic to HTS. Furthermore, recent
penetration depth measurements in LSCO and HgBa$_2$CuO$_{4+\delta}$ showed that
the $c$-axis zero-temperature superfluid density $\lambda_{c}^{-2}(0)$ exhibits similar
behaviour \cite{CP4}. The apparent competition between quasi-static magnetic
correlations and superconductivity thus results in a crossover to weak
superconductivity characterised by a strong suppression of the superfluid density. This
suppression in the underdoped region can also be directly linked to the strong reduction
in entropy and condensation energy associated with the pseudogap \cite{Loram}. It is
therefore evident here that the onset of short-range magnetic correlations at the critical
point $p_c$ coincides with an abrupt change in the superconducting ground state
properties in HTS. It separates the phase diagram into two distinct regions: (i) below
$p_c$ where $T_g$, $T_f$ increase rapidly with underdoping and the superfluid density
is rapidly suppressed, and (ii) above $p_c$ where $T_g$, $T_f$ =0 and the superfluid
density is almost constant, indicating a transition from weak to strong superconductivity,
respectively. Moreover, it is at $p_c$ where other fundamental properties such as the
superconducting condensation energy and quasiparticle life-time, change abruptly, the
resistivity follows its unusual linear temperature dependence to the lowest temperature
and the pseudogap extrapolates to zero \cite{TallonLoram,Loram}. These features all
indicate that the quantum transition identified here is connected with the fundamental
physical properties of HTS.

In general terms our results complement a growing body of work pointing to an intimate
relation between slow magnetic correlations and superconductivity that seems mutually
co-operative in some experiments \cite{Kimura,Fujita,Lake} and competitive in others
(present work and ref. \cite{Singer}. In the superconducting state an energy gap is
observed in the spin spectrum \cite{Yamada} but its $d$-wave nature ensures that low
energy spin fluctuations may still be present, as observed here. Inelastic neutron
scattering \cite{Kastner,Kimura} and NMR \cite{Julien} experiments show that Zn
substitution slows spin fluctuations and suppresses long range order. They are consistent
with our observation that Zn doping enhances spin glass behaviour. These studies provide
a context for the present work but the key new result here is the observation of a quantum
transition at a critical doping with an associated change in the superconducting ground
state. It follows from our work that the elusive physics of HTS may reduce to the known
generic physics of materials near a quantum transition, thus explaining many of their
unconventional properties \cite{SC,SY1,CC,CV1,DP,Si}.

In conclusion we have performed a comprehensive study of low-energy spin fluctuations
and of the superfluid density in several HTS. We found that both low-energy spin
fluctuations and the spin-glass state disappear abruptly at a critical doping $p_c$ at zero
temperature. Moreover, the rate of slowing of these fluctuations is found to diverge at
this point. This provides the first evidence for a $quantum$ $glass$ $transition$ in HTS
and indicates quantum critical fluctuations and associated dynamical crossovers may
dominate the essential physics of high temperature superconductivity. The identified
critical point shows that there are two distinct ground states in the superconducting
phase, manifested in a crossover from strong superconductivity for $p>p_c$ to weak
superconductivity for $p<p_c$. The identified transition bears a close resemblance to the
locally-critical two dimensional quantum phase transitions described recently by Si $et$
$al$ \cite{Si} in which the magnetic correlations are localised in space but have
unlimited range in time.

We are grateful to A.D. Taylor of the ISIS Facility, Rutherford Appleton Laboratory for
the allocation of muon beam time. Thanks are due to B. Ingham for synthesising some of
the Zn-doped samples. C.P. thanks C. Bernhard, J.I. Budnick, S. Chakravarty, A.
Chubukov, S.A. Kivelson W.Y. Liang, J.W. Loram, A.J. Millis, Ch. Niedermayer, D.
Pines, S. Sachdev, J. Schmalian and C.M. Varma for useful discussions and The Royal
Society for a University Fellowship. J.L.T. acknowledges financial assistance from the
New Zealand Marsden Fund and T.X. from the National Natural Science Foundation of
China.

$Correspondence$ $should$ $be$ $addressed$ $to$ $C.P.$ (cp200@hermes.cam.ac.uk).

FIGURE CAPTIONS

FIG. 1.  Typical zero-field $\mu $SR spectra of La$_{2-x}$Sr$_x$CuO$_4$ for
$Sr=0.08$ measured at (a) 1.3K and (b) 29K. The solid line in (b) is a fit to a stretched
exponential, discussed in the text. In (a) the solid line is a fit to a Lorentzian, describing
the initial rapid drop of asymmetry, plus a stretched exponential, describing the long-time
tail.

FIG. 2.  The temperature dependence of (a) the (stretched-exponential) exponent $\beta$
and (b) the relaxation rate, $\gamma$, within the Kubo-Toyabe function obtained by
fitting muon depolarisation data for La$_{2-x}$Sr$_x$CuO$_4$ ($Sr$=0.08).

FIG. 3.  The doping dependence of the crossover temperatures $T_f$  (open symbols)
where the spin fluctuations enter the $\mu$SR time window and $T_g$ (closed symbols)
where the spin fluctuations leave the $\mu$SR time window. Below $T_g$ the
fluctuations freeze out into a spin glass. Black, red and blue data are for La$_{2-
x}$Sr$_x$Cu$_{1-y}$Zn$_y$O$_4$ with $y=0$, $y=0.01$ and $y=0.02$, respectively.
The green symbols are $T_f$ (open) and $T_g$ (closed) values, determined in the same
way, for Bi$_{2.1}$Sr$_{1.9}$Ca$_{1-x}$Y$_x$Cu$_2$O$_{8+y}$. The $T_c$ values
for all samples are shown as crosses in the respective colours. $T_c$ for
Bi-2212 have been divided by 2 for clarity.

FIG. 4.  Doping dependence of the inverse square of the zero temperature in-plane
penetration depth for pure ($y$=0) La$_{2-x}$Sr$_x$CuO$_4$ measured by the $ac$-
susceptibility technique.

\end{multicols}

\end{document}